\documentclass[aps,pre,reprint,superscriptaddress,showpacs,showkeys]{revtex4-1}

\usepackage{amsmath}
\usepackage[usenames]{color}
\usepackage{amssymb}
\usepackage{graphicx}

\DeclareMathOperator{\e}{e}

\begin{document}

\title{Experimental control of chaos by variable and distributed delay feedback}

\author{Thomas J\"ungling}
\email{thomas.juengling@physik.uni-wuerzburg.de}
\affiliation{Institute for Theoretical Physics, University of W\"urzburg, Am Hubland, D-97074, W\"urzburg, Germany}

\author{Aleksandar Gjurchinovski}
\affiliation{Institute of Physics, Faculty of Natural Sciences and Mathematics, Saints Cyril and Methodius University, P.\ O.\ Box 162, 1000 Skopje, Macedonia}

\author{Viktor Urumov}
\affiliation{Partenij Zografski 46, 1000 Skopje, Macedonia}

\begin{abstract}
We report on a significant improvement of the classical time-delayed feedback control method for stabilization of unstable periodic orbits or steady states. In an electronic circuit experiment we were able to realize time-varying and distributed delays in the control force leading to successful control for large parameter sets including large time delays. The presented technique makes advanced use of the natural torsion of the orbits, which is also necessary for the original control method to work.
\end{abstract}


\date{\today}

\maketitle

\section{Introduction}
\label{introduction}

The control of chaos has attracted the attention of physicists since over two decades now. The most established method for controlling unstable periodic orbits embedded in a chaotic attractor is the application of a continuous time-delayed feedback, which was first introduced by Pyragas \cite{Pyragas:92}. A dynamical system
\begin{equation}
\dot{\mathbf{x}}=\mathbf{f}(\mathbf{x})
\label{dynsys}
\end{equation}
subjected to Pyragas control becomes
\begin{equation}
\dot{\mathbf{x}}=\mathbf{F}\left(\mathbf{x},k(\mathbf{x}_\tau-\mathbf{x})\right)\;,
\label{dynsyscontr}
\end{equation}
where $\mathbf{x}_\tau\equiv\mathbf{x}(t-\tau)$, and $k$ is the feedback gain. The control is applied in a way that $\mathbf{F}(\mathbf{x},\boldsymbol{0})=\mathbf{f}(\mathbf{x})$. To control an unstable periodic orbit, the delay $\tau$ is adjusted to match the period of the orbit. When stabilization is achieved, the control signal vanishes by construction, meaning that the controlled orbit remains a solution of the original Eq.~(\ref{dynsys}). This property of the Pyragas method is known as noninvasiveness. For stabilization of unstable steady states, the choice of the delay time $\tau$ is not as restrictive as in the case of unstable periodic orbits, and the interval of $\tau$ for which the control is successful is shown to be system-dependent.

The Pyragas method was successfully implemented in a variety of experimental setups, and an effort has been put into progress to generalize or modify the original control scheme in order to improve its performance. An overview of the field is given in \cite{Schuster:07}. An important generalization of the Pyragas method was proposed by Socolar et al. \cite{Socolar:94}, where the feedback signal was taken in the form of a geometric sum (extended time-delayed feedback control, ETDFC), or a mean value of a finite number of delay terms \cite{Socolar:98} (N time-delayed feedback control, NTDFC), each using information from many previous states of the system involving integer multiples of the delay time $\tau$. A natural extension of this method was proposed by Ahlborn and Parlitz \cite{Ahlborn:04}, using two or more delayed feedback signals with incommensurate delay times. These multiple delay extensions achieve stabilization of unstable states with a higher degree of instability.

In a recent work \cite{Gjurchinovski:08}, it was shown that the efficiency of the Pyragas method to control unstable steady states is significantly improved by applying a variable time-delay in the original control scheme. We call this method variable-delay feedback control (VDFC). In this case, the modulation of the delay is in a fixed interval around some nominal delay value, and it can be realized in a deterministic or a stochastic way. The essence of the improved control mechanism lies in the delay \textit{distribution}, which is created by the modulation. It has already been reported, that delay distributions in coupled oscillators lead to stabilization of steady states \cite{Atay:03,Konishi:10,Kyrychko:11}. The same mechanism applies to self-feedback on a single oscillator.

In this work we experimentally investigate the control of steady states and periodic orbits using variable-delay feedback as well as distributed-delay feedback. In Sec.~\ref{experiment} the experimental setup is described. The possibilities and constraints for the construction of variable and distributed delays are shown. The relation between modulation and distribution is explained in Sec.~\ref{moddist}. In Sec.~\ref{fixedpoint} we study three different methods of fixed point control, whereas Sec.~\ref{orbits} deals with the control of a periodic orbit. Finally in Sec.~\ref{mechanism} the control mechanism, which is mostly covered theoretically, is summarized and interpreted. The last section contains our conclusions.

\section{Experiment}
\label{experiment}

\begin{figure}
\includegraphics[width=0.7\columnwidth]{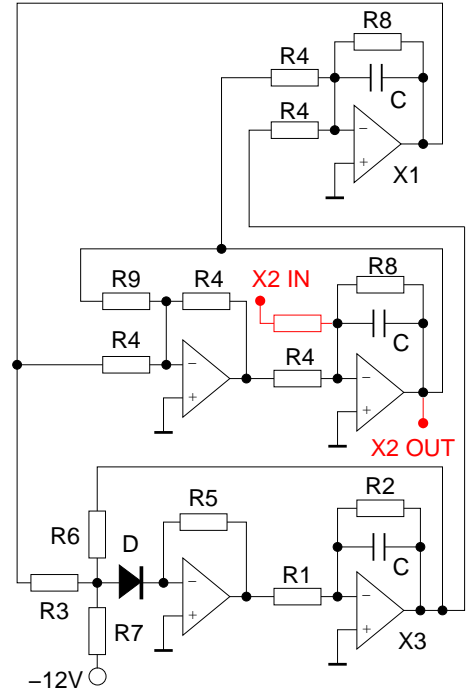}
\caption{The circuit scheme of the autonomous diode oscillator with inputs and outputs for coupling. Values of components: $R_1=2.7\mathrm{k\Omega}, R_2=3.6\mathrm{k\Omega}, R_3=\mathrm{7.5k\Omega}, R_4=\mathrm{10k\Omega}, R_5=13\mathrm{k\Omega}, R_6=15\mathrm{k\Omega}, R_7=33\mathrm{k\Omega}, R_8=200\mathrm{k\Omega}$. $R_9$ is variable $0-50\mathrm{k\Omega}$ and determines the system parameter $a$ according to $R_9=\mathrm{10k\Omega}/a$ (accuracy of resistors 1\%). Capacitators: $C=10\mathrm{nF}$ (accuracy 5\%). Type of diode: 1N4007 from DC Components. Type of operational amplifiers: TL084.}
\label{schema}
\end{figure}

Our experimental setup consists of a single electronic oscillator subjected to a time-delayed feedback. A schematic of the circuit is shown in Fig.~\ref{schema}. This circuit has already been used in different contexts and is described there in detail \cite{Juengling:10,Juengling:11,Heiligenthal:11}. The equations of motion for the main dynamical components are similar to the R\"ossler system and read explicitly
\begin{equation}
\begin{split}
\dot{x}&=\omega_0(-cx-y-z)\\
\dot{y}&=\omega_0(x+(a-c)y)\\
\dot{z}&=\omega_0(g(x,z)-\gamma z)\;,
\end{split}
\label{circuitdgl}
\end{equation}
where $g(x,z)=b(|w|+w)$ with $w=x+z/2-z_{thr}$. The variables $x,y,z$ are voltages taken at X1, X2, and X3, respectively. The timescale of the oscillator is determined by $\omega_0=1/R_4C=10\text{ks}^{-1}$. The parameters $c=0.05,b=3.18,\gamma=2.82,$ and $z_{thr}=3.35$V are optimized in a way to obtain robust dynamics and a well-defined bifurcation structure by variation of the main circuit parameter $a$. The standard value used in most experiments is $a=0.3$, for which a chaotic attractor with a maximum Lyapunov exponent of $\Lambda=0.1\omega_0$ is observed.

\begin{figure}
\begin{center}
\includegraphics[width=0.8\columnwidth]{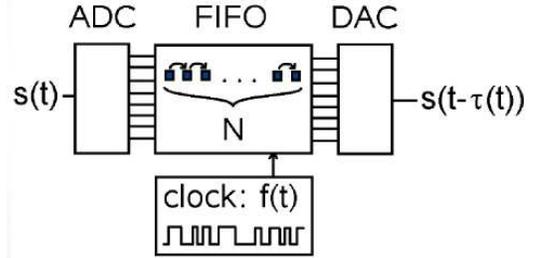}
\end{center}
\caption{A single delay line with clock frequency modulation.}
\label{delayline}
\end{figure}

We apply the time-delayed feedback to the $y$-component, whose governing equation is thus changed into
\begin{equation}
\dot{y}=\omega_0\left(x+(a-c)y+k(y_\tau-y)\right),
\end{equation}
where $y_\tau=y(t-\tau)$ is the delayed $y$-signal, and $k$ is the feedback gain parameter. In order to get a clear time-delayed signal of this component with delay times $\tau$ in the order of miliseconds, we use digital delay lines as shown in Fig.~\ref{delayline}. The analog signal is first prepared to fit in the range between $0$V and $5$V. Then it is discretized by an analog-to-digital converter (ADC) with 8 bit. Delay is created by storing the signal in a first-in-first-out device (FIFO) with a maximum of 1kByte memory. The device is triggered by an externally prepared write-clock and read-clock, which defines the number $N$ of stored samples of the signal and the rate $f$, by which they are shifted through the register. Finally the output of the FIFO is converted back by a digital-to-analog converter (DAC) and low-pass filtered to clear the step-like behavior of the signal, before rescaling to the original voltage range. The low-pass filter is of second order with a cutoff frequency $f_{lp}=20$kHz. The freqency gap between $f_{lp}$ and the sampling frequency $f$ is used for the dithering effect by applying a suitable small noise signal, which interpolates between the bits of the ADC. Under optimal constraints we can effectively reach a resolution of 11bit instead of 8bit provided by the devices without dithering. We have constructed several such delay lines, which can be adjusted to different delays simultaneously so we can implement a large variety of different feedback signals.

The clock signal for the delay line is created by a function generator type Agilent 33220A. It allows an arbitrary modulation of the clock frequency, which results in a delay time modulation $\tau(t)$. The relation between clock frequency modulation and delay time modulation for the given delay lines can be derived as
\begin{equation}
N=\int_{t-\tau(t)}^t f(t')dt'\;.
\label{Nft}
\end{equation}
Here $N$ is the buffer size of the FIFO, $f(t)$ is the current clock frequency and $\tau(t)$ the current delay time, i.e., if $y(t)$ is the signal at the input of the delay line, then we have at the output $y(t-\tau(t))$. Eq.~(\ref{Nft}) does not allow for an arbitrary delay time modulation, for instance we cannot realize a jump in $\tau(t)$. For a delay modulation which includes step-like functions $\tau(t)$ we therefore choose to switch between different delay lines. A similar technique is applied in Sec.~\ref{orbits}, where only discrete values of the delay time are allowed.

\section{Modulation and distribution}
\label{moddist}

In the following, we consider the types of modulation for which we can write
\begin{equation}
\tau(t)=\tau_0+\varepsilon\tilde{\tau}(t)\;,
\end{equation}
where $\varepsilon$ is the modulation amplitude and $\tilde{\tau}(t)\in[-1,1]$ is a periodic signal with period $T_m=f_m^{-1}$. It may also be an irregular bounded signal, for which the power spectrum has at least a lower cutoff frequency $f_{co}$. In the limit case of a very high modulation frequency $f_m$ or a very high cutoff frequency $f_{co}$, respectively, the modulated time delay is equivalent to a delay distribution $\rho(\tau)$. The density function $\rho(\theta)$ is defined in a way that for an increment $d\theta$ the fraction of time  that $\tau(t)$ is found to lie between $\theta$ and $\theta+d\theta$ is given by $\rho(\theta)d\theta$. A rigorous result establishing the relation between variable delay and the corresponding delay distribution was reported by Michiels et al. \cite{Michiels:05}. An example is provided by rectangular (square wave) modulation, in which $\tau(t)$ switches periodically between $\tau_0-\varepsilon$ and $\tau_0+\varepsilon$, both for the same duration. This leads to a two-peak distribution
\begin{equation}
\rho(\theta)=\tfrac{1}{2} \left(\delta(\theta-\tau_0+\varepsilon)+\delta(\theta-\tau_0-\varepsilon)\right)\;.
\label{moddistex1}
\end{equation}
Another example is a triangular (sawtooth) modulation. The delay $\tau(t)$ rises and falls linearly between its extreme values, which results in a uniform distribution. It is not affected by the skewness of the sawtooth wave and we always obtain
\begin{equation}
\rho(\theta)=
\begin{cases}
\frac{1}{2\varepsilon} & ,\;\theta\in[\tau_0-\varepsilon,\tau_0+\varepsilon],\\
0 & ,\;\text{else.}
\end{cases}
\label{moddistex2}
\end{equation}
Finally, a sinusoidal modulation leads to
\begin{equation}
\rho(\theta)=\frac{1}{\pi\sqrt{\varepsilon^2-(\theta-\tau_0)^2}}\;.
\label{moddistex3}
\end {equation}
As we show later, in most cases in which delay time modulation improves stability, the mechanism can essentially be reduced to the delay distribution created by the modulation. However, there are also cases, in which a finite modulation frequency plays a role for stabilization. This issue is discussed in Sec.~\ref{orbits}.

\section{Stabilization of fixed points}
\label{fixedpoint}

As we have shown previously in a theoretical work \cite{Gjurchinovski:08}, a delay time modulation may enhance the stability properties of a Pyragas-controlled fixed point significantly. We demonstrate this effect in our electronic experiment using different types of delay modulations (distributions). The free-running circuit has an unstable fixed point at the coordinate origin. Eqs.~(\ref{circuitdgl}) also have another fixed point solution, but it lies outside the experimental voltage range, so we investigate only the central fixed point $(x_0,y_0,z_0)=(0,0,0)$. Its stability is determined by the eigenvalues of $\mathbf{Df}(\boldsymbol{0})$, which denotes the Jacobian of the vector field $\mathbf{f}$ at the fixed point
\begin{equation}
\mathbf{Df}(\boldsymbol{0})=\omega_0
\begin{pmatrix}
-c & -1 & -1 \\
1 & a-c & 0 \\
0 & 0 & -\gamma
\end{pmatrix}\;.
\end{equation}
The eigenvalues of this matrix are $\lambda_{1/2}=\omega_0\left(\tfrac{a}{2}-c\pm\sqrt{a^2/4-1}\right)$ and $\lambda_3=-\gamma\omega_0$. For standard parameters the first two are a complex conjugate pair with positive real part, $\lambda_{1/2}\approx(0.1\pm i)\omega_0$, and the third one is negative, $\lambda_{3}=-2.82\omega_0$. The fixed point is susceptible to Pyragas control, because the imaginary parts of $\lambda_{1/2}$ are nonzero, so that the neighborhood of the fixed point undergoes a $2\pi$ torsion in the time interval $T_0\approx 2\pi/\omega_0=0.63$ms. For the sake of simplicity we further discuss the stability properties of the fixed point in terms of a normal form
\begin{equation}
\dot{z}=\lambda_0 z
\label{fpnocontrol}
\end{equation}
with $z\in\mathbb{C}$ and $\lambda_0=(a/2-c)+i$. The fixed point is at the origin $z=0$ and the oscillation period around it is $T_0=2\pi$. The results obtained by the normal form are in very good agreement with all our findings from the full system of differential equations as well as with the experimental outcomes. A Pyragas-type feedback changes Eq.~(\ref{fpnocontrol}) to
\begin{equation}
\dot{z}=\lambda_0 z+\kappa(z_\tau-z)\;,
\label{nform}
\end{equation}
with $z_\tau\equiv z(t-\tau)$. To identify the normalized coupling $\kappa$ with the experimental feedback gain, we have to set $\kappa=k/2$. The corresponding characteristic equation states
\begin{equation}
\lambda=\lambda_0+\kappa\left(e^{-\lambda\tau}-1\right),
\label{char}
\end{equation}
which follows by applying the usual ansatz $z\sim e^{\lambda t }$ in Eq. (\ref{nform}).
Control is in principle successful if $\Re(\lambda_0)\tau<2$ \cite{Just:99}. Within this constraint, the delay times for which control is most effective are odd multiples of $T_0/2$, and control is not possible for even multiples of $T_0/2$. 
If the fixed point is stabilized, the control force $\kappa(z_\tau-z)$ vanishes, so the control is non-invasive. A delay-time modulation (or the use of distributed delays) will not change this feature, since at the fixed point $z_\tau\equiv z$ for all values of $\tau$, provided that the control is successful. Therefore we can apply any delay modulation (distribution) and study its effect on the stability of the fixed point. Within our experimental constraints we investigate three different types of variable (distributed) delay for control.

\subsection{Clock frequency modulation}

\begin{figure}
\includegraphics[width=0.7\columnwidth]{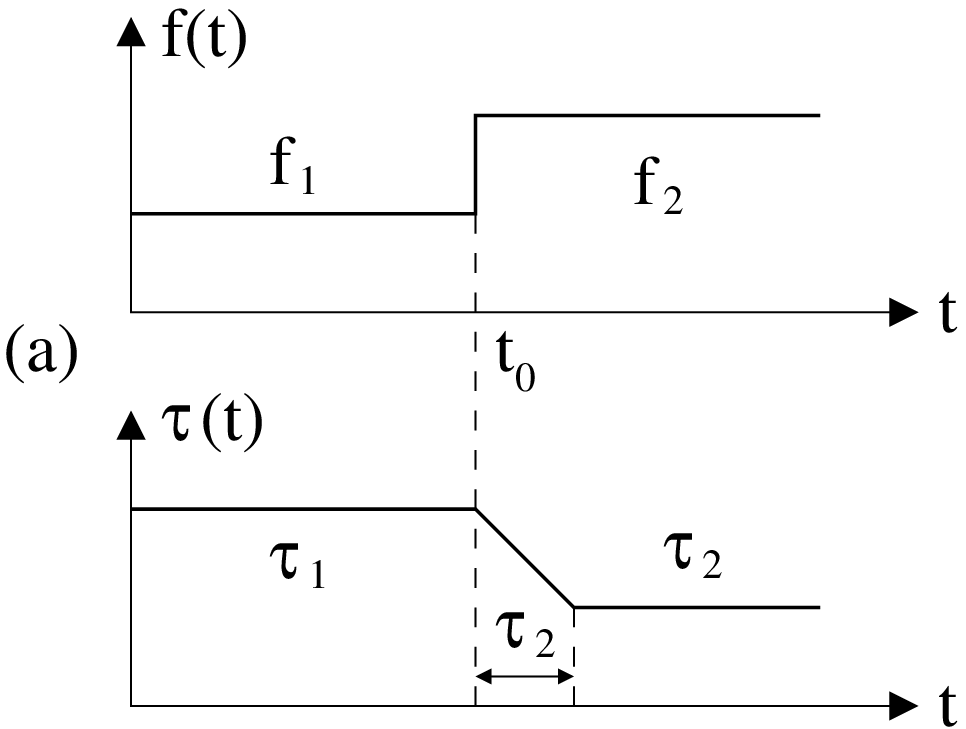}\\[3ex]
\includegraphics[width=0.7\columnwidth]{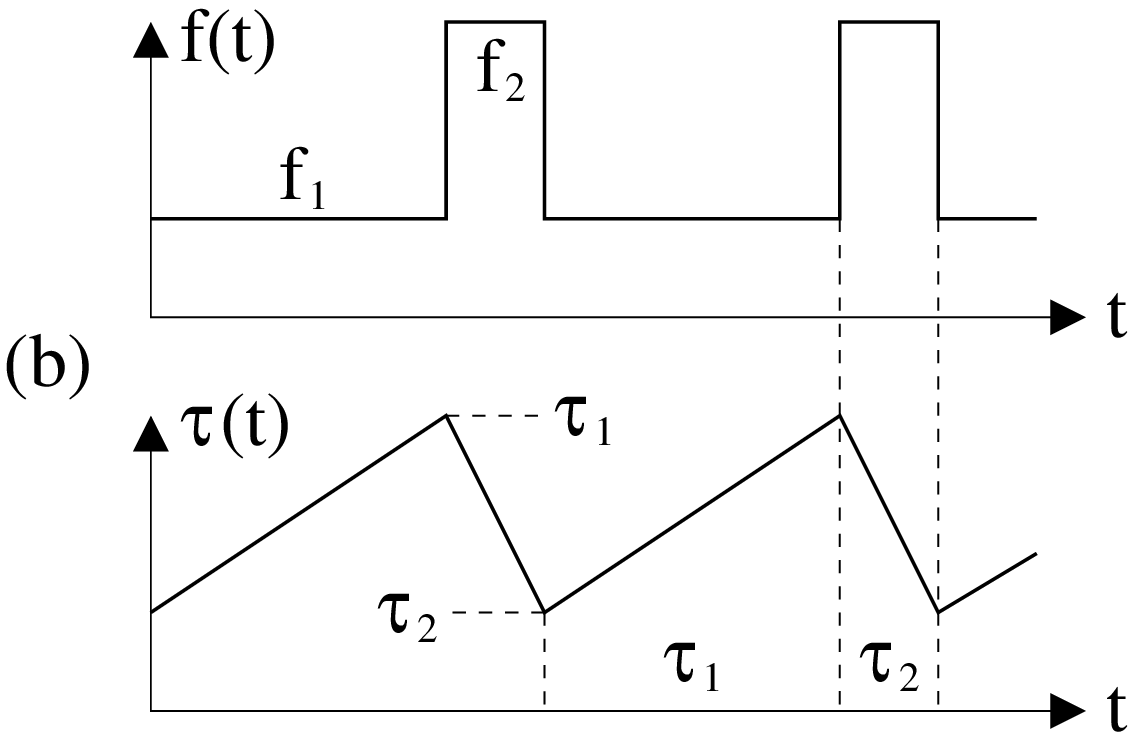}
\caption{Example relations between clock frequency modulation $f(t)$ and delay time modulation $\tau(t)$ after Eq.~(\ref{Nft}). a) Effect of an instantaneous increase in clock freqency. b) Construction of a sawtooth wave.}
\label{modexamples}
\end{figure}
Delay modulation is created via modulation of the clock frequency, which drives the delay lines. All possible relations between $f(t)$ and $\tau(t)$ are given by Eq.~(\ref{Nft}). Fig.~\ref{modexamples} shows two examples, which satisfy this equation. Example~(a) demonstrates the effect of a step function
\begin{equation}
f(t)=
\begin{cases}
f_1 & ,\; t<t_0,\\
f_2 & ,\; t\ge t_0,
\end{cases}\;
\end{equation}
with $f_2>f_1$. In this case $\tau(t)$ follows a kink function, which at $t=t_0$ starts changing from the left constant value $\tau_1=N/f_1$ to the right constant value $\tau_2=N/f_2$. Both are trivial cases of a non-modulated clock frequency. The duration of the transition is given by $\tau_2$, because this is the time, that the buffer - which was previously recorded with rate $f_1$ - needs to run through the FIFO with the new rate $f_2$.

\begin{figure}
a)\includegraphics[width=0.9\columnwidth]{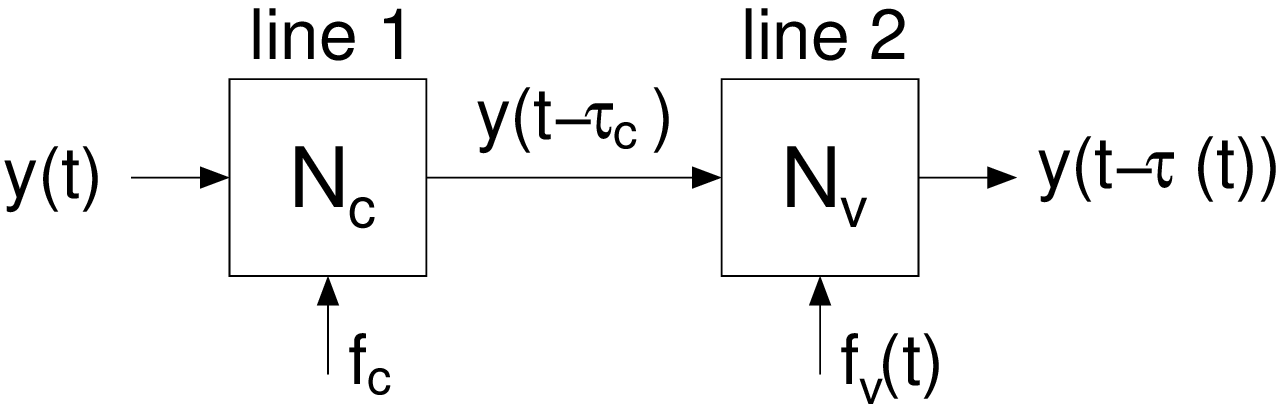}\\[2ex]
b)\includegraphics[width=0.9\columnwidth]{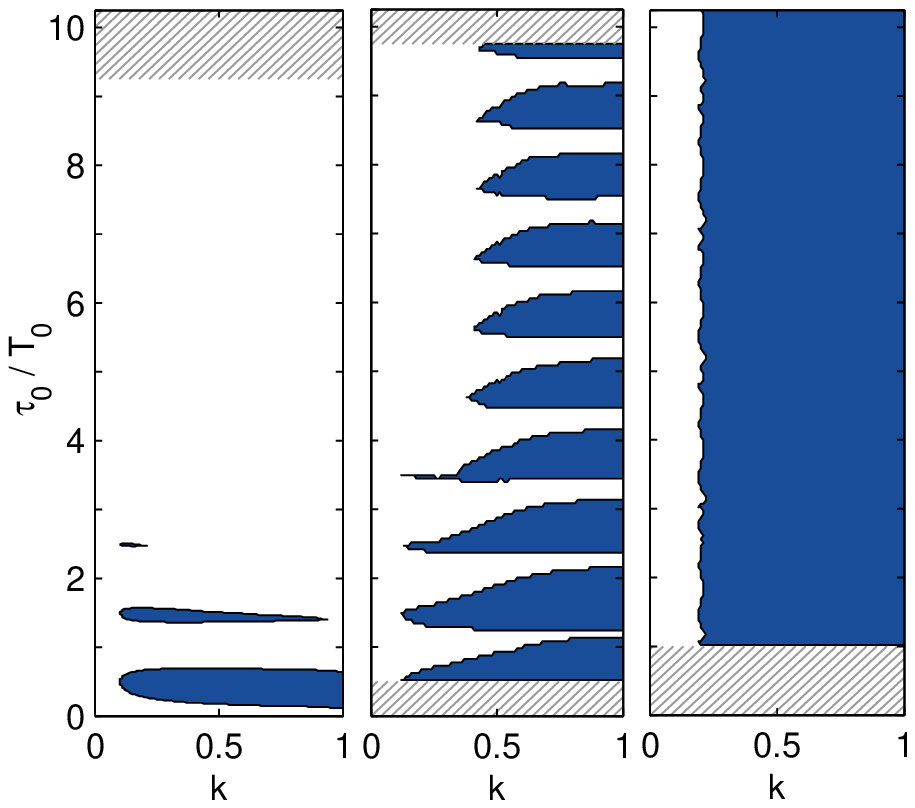}
\caption{Delay modulation for fixed point control. a) Experimental setup with two delay lines for a modulation $\tau(t)$ between $\tau_0-\varepsilon$ and $\tau_0+\varepsilon$. b) Scan of parameters $k$ and $\tau_0$ for different values of modulation amplitude $\varepsilon$. White: No control, oscillations with $\sigma_y>0.1$V. Blue: Control domains. Hatched areas mark parameter constellations that are not examined. Left: $2\varepsilon=0$ (time-delayed feedback control). Middle: $2\varepsilon=T_0/2$ with $T_m=T_0$. Right: $2\varepsilon=T_0$ and $T_m=2T_0$. Oscillation period $T_0=2\pi/\omega_0$.}
\label{FPmod123}
\end{figure}

Such a ramp is helpful in creating a triangle modulation in $\tau(t)$, which corresponds to a uniform delay distribution between two extremum values like in Eq.~(\ref{moddistex2}). The clock frequency modulation fulfilling the restrictions to create a triangle-wave is an asymmetric rectangular wave with
\begin{equation}
f(t)=
\begin{cases}
f_1 & \text{, duration } \tau_1,\\
f_2 & \text{, duration } \tau_2\;.
\end{cases}
\end{equation}
Here $T_m=\tau_1+\tau_2$, $\tau_1=N/f_1$, and $\tau_2=N/f_2$. Fig.~\ref{modexamples}b illustrates the creation of such a modulation. If the intervals in $f(t)$ were not set correctly, the sawtooth in $\tau(t)$ would show clipping effects. Note that a similar modulation could also be achieved using larger values of $N$ with the same frequency modulation, because the triangular pattern reappears periodically by increasing $N$. But this also involves a higher mean delay time, which is not desirable, because the possible values are fixed by the choice of the modulation amplitude. So we restrict our setup to the `first Brillouin zone' in $N$. Returning to our previous notation using $\tau_0$ and $\varepsilon$ we now aim to change those parameters independently. To this end we use two delay lines serially. The first delay line is adjusted to a constant delay and the second line receives the described clock frequency modulation. The buffer sizes of both lines can be adjusted separately, so we have control over both mean delay time $\tau_0$ and modulation amplitude $\varepsilon$. Fig.~\ref{FPmod123}a shows the experimental setup for delay modulation. To keep the modulated clock frequency in a useful range, we choose $f_{v.1}=810$kHz and $f_{v.2}=270\text{kHz}=f_{v.1}/3$. For a given modulation amplitude $\varepsilon$ we then set the buffer size $N_v=f_{v.1}\cdot\varepsilon$, so that $\tau_1=\varepsilon$ and $\tau_2=3\varepsilon$. The modulation period is $T_m=4\varepsilon$. By this choice an offset $2\varepsilon$ occurs in $\tau_0$, i.e., if we set the delay of the first line to $\tau_c=0$, then we still have a mean delay $\tau_0=2\varepsilon$. For a fixed modulation amplitude we scan the parameter space of feedback gain $k$ and mean delay time $\tau_0$ automatically by setting $N_c$ and $f_c$. For each constellation the oscillation strength of the circuit is measured in terms of the standard deviation $\sigma_y$ of the signal $y(t)$. If the oscillation is below some pre-defined value, here $y_{thr}=0.1$V, we define the dynamics as a stable fixed point. The results are a visualization of the control domains as shown in Fig.~\ref{FPmod123}b.
Without delay modulation ($\varepsilon=0$) we recognize the stability islands, which are typical for TDFC. Stabilization of the fixed point is sensitive to the phase $\omega_0\tau$ and limited by $\Re\{\lambda_{1/2}\}\tau<2$, which in our case yields $\tau_0<3.2T_0$. Here $T_0=2\pi/\omega_0=0.63$ms is the natural period of the uncontrolled oscillator. With increasing modulation amplitude more and more of the parameter plane is covered by stable solutions and the limitation in delay time is overcome. Setting the modulation amplitude equal to $\varepsilon=T_0/2$ appears to have the strongest effect (right panel of Fig.~\ref{FPmod123}b), although the increase of modulation amplitude has come along with a decrease of the modulation frequency. The results match the theoretical prediction for the distributed delay limit very well, so we conclude that the slow modulation does not significantly impede the mechanism, which has been found for high modulation frequencies.

\subsection{Low-pass filter}

\begin{figure}
\includegraphics[width=\columnwidth]{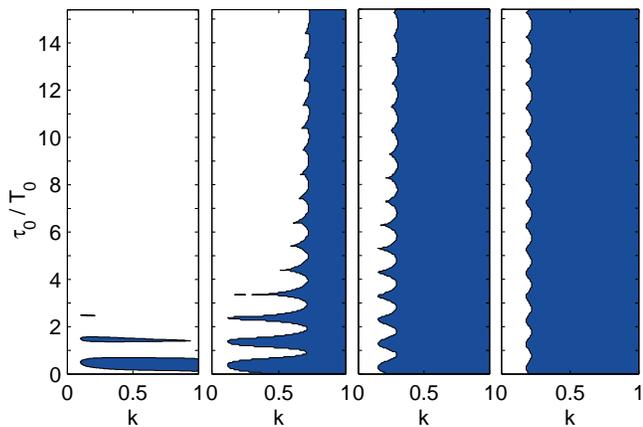}
\caption{Effect of a delay distribution created by a low-pass filter with filter constant $\beta$. Signal before the filter has a delay $\tau_0$. Control domains ($\sigma_y<0.1$V) marked in blue. Left: $\beta\rightarrow\infty$, direct signal, TDFC. Left middle: $\beta=\omega_0$. Right middle: $\beta=\omega_0/3$. Right: $\beta=\omega_0/10$.}
\label{fptpf123}
\end{figure}
A rather trivial way to create a special delay distribution is the use of a low-pass filter. Let this filter be defined by
\begin{equation}
\dot{w}=\beta(s(t)-w)\;,
\label{lpf}
\end{equation}
in which $w\in\mathbb{R}$ and $\beta>0$. The external signal $s(t)$ is low-pass filtered, which can be seen in the integral form of Eq.~(\ref{lpf})
\begin{equation}
w(t)=\int_{-\infty}^t s(t')\beta\e^{-\beta(t-t')}dt'\;,
\end{equation}
which leads to the spectral power density
\begin{equation}
S_w(f)=\frac{S_s(f)}{1+\frac{4\pi^2f^2}{\beta^2}}\;.
\end{equation}
So if we low-pass filter the output of a delay line, before further constructing the control signal, the equations of motion of the controlled circuit turn to
\begin{equation}
\begin{split}
\dot{x}&=\omega_0\left(-cx-y-z\right)\\
\dot{y}&=\omega_0\left(x+(a-c)y+k(w-y)\right)\\
\dot{z}&=\omega_0\left(g(x,z)-\gamma z\right)\\
\dot{w}&=\beta(y_{\tau_0}-w)
\end{split}
\end{equation}
and we realize a distributed delay feedback with the weight function
\begin{equation}
\rho(\theta)=
\begin{cases}
0 & ,\; \theta < \tau_0, \\
\beta\e^{-\beta(\theta-\tau_0)} & ,\; \theta \ge \tau_0\;.
\end{cases}
\end{equation}
Note that $\tau_0$ here is not the mean delay time of the distribution. The measurement is performed analogous to the previous one, here for different values of $\beta$, which determines the distribution width. Low values of $\beta$ correspond to a broad distribution due to a slowly reacting filter, and for $\beta\rightarrow\infty$ the filter simply transmits the signal of the delay line without change. The resulting control domains for this method are shown in Fig.~\ref{fptpf123}, and we recognize a similar behaviour as for the modulation case. For the broader distributions we did not observe any limitation of the stabilization at large delay times.

\subsection{Two delays}

\begin{figure}
a)\includegraphics[width=0.9\columnwidth]{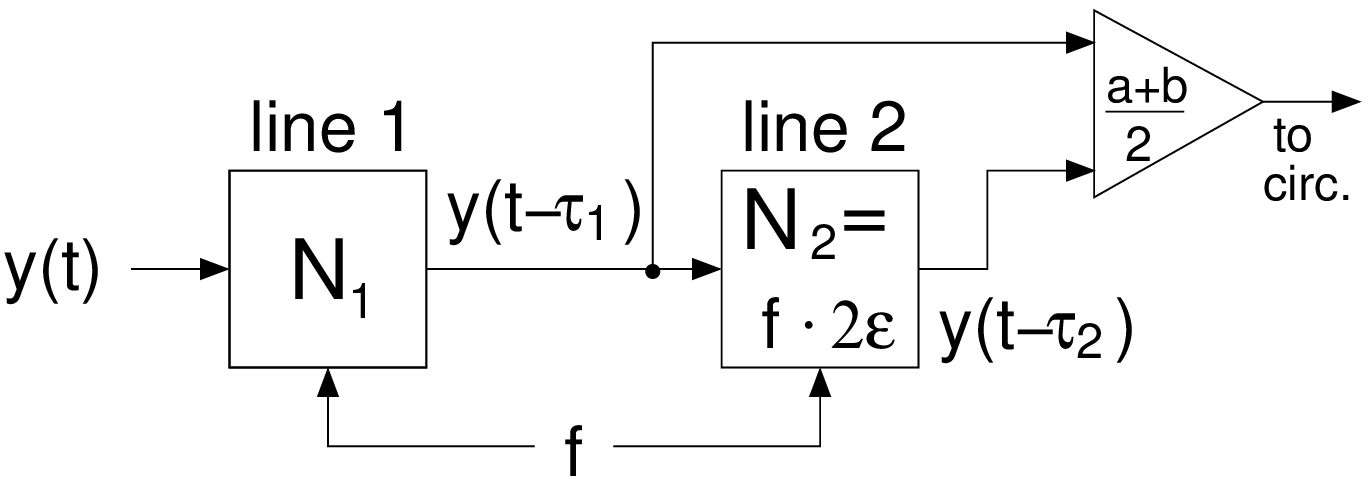}\\[2ex]
b)\includegraphics[width=\columnwidth]{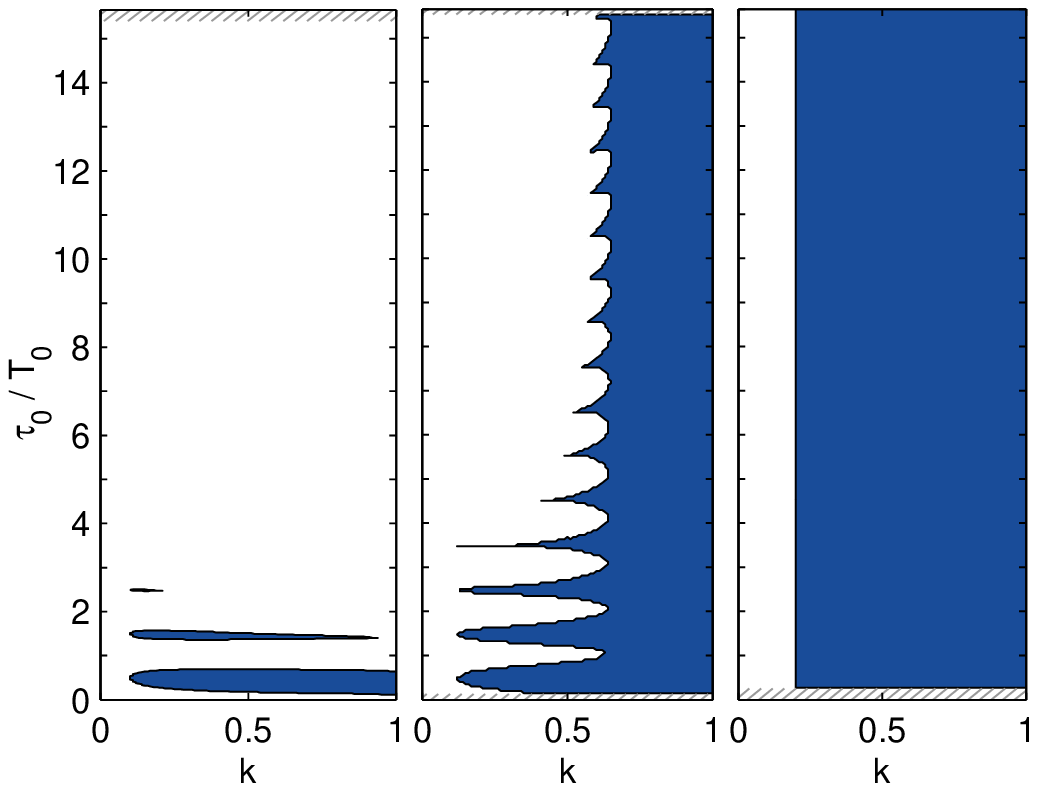}
\caption{Superposition of two delays. a) Experimental setup. Delay of the first line is $\tau_1$. The delay of the second line is kept constant on $\tau_2-\tau_1=2\varepsilon$. b) Regions of fixed point control (blue, $\sigma_y<0.1$V). Hatched areas are excluded from measurement. Left: $2\varepsilon=0$ (time-delayed feedback control). Middle: $2\varepsilon=T_0/4$. Right: $2\varepsilon=T_0/2$.}
\label{fptwodel123}
\end{figure}
We repeat our parameter scan using two delay lines as shown in Fig.~\ref{fptwodel123}a. The signals are superposed in a way, that the feedback force becomes
\begin{equation}
k\left(\tfrac{1}{2}(y_{\tau_1}+y_{\tau_2})-y\right)\;.
\label{twodeleq}
\end{equation}
The delay of the first line is $\tau_1=\tau_0-\varepsilon$ and the delay of the second line is $\tau_2=\tau_0+\varepsilon$. This setup corresponds to a modulation of the delay time, in which the delay switches between $\tau_1$ and $\tau_2$ with a high frequency. We vary $k$ and $\tau_0$ for some selected values of $\varepsilon$. Again, by the oscillation strength $\sigma_y$ of $y(t)$ we determine, whether fixed point stabilization is possible. The results are shown in Fig.~\ref{fptwodel123}b and they correspond well to the previous findings. The optimal value of delay difference is $2\varepsilon=T_0/2$. For this case also no limitation for large delays was observed. The possibility to control the unstable fixed point at arbitrarily large delay times $\tau_0$ is a striking feature, which all the investigated modulation and distribution techniques have in common.

\subsection{Robustness of the control methods}

\begin{figure}
\includegraphics[width=\columnwidth]{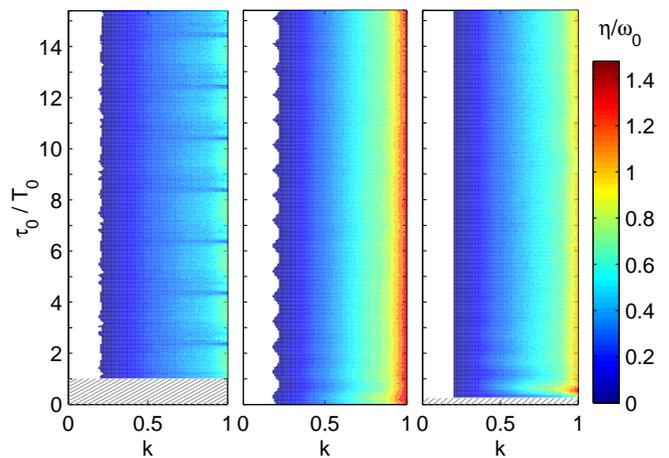}
\caption{Parameter scans with additive noise. White: No control ($\sigma_y>0.1$V). Colors correspond to $\eta(k,\tau_0)$ in the control domains. Hatched: Not measured. Left: Control by variable-delay feedback with $2\varepsilon=T_0$ and $T_m=2T_0$. Middle: Low-pass filtered feedback with $\beta=\omega_0/10$. Right: Two-delay feedback with $2\varepsilon=T_0/2$.}
\label{noisyscans}
\end{figure}

\begin{figure}
a)\includegraphics[width=0.9\columnwidth]{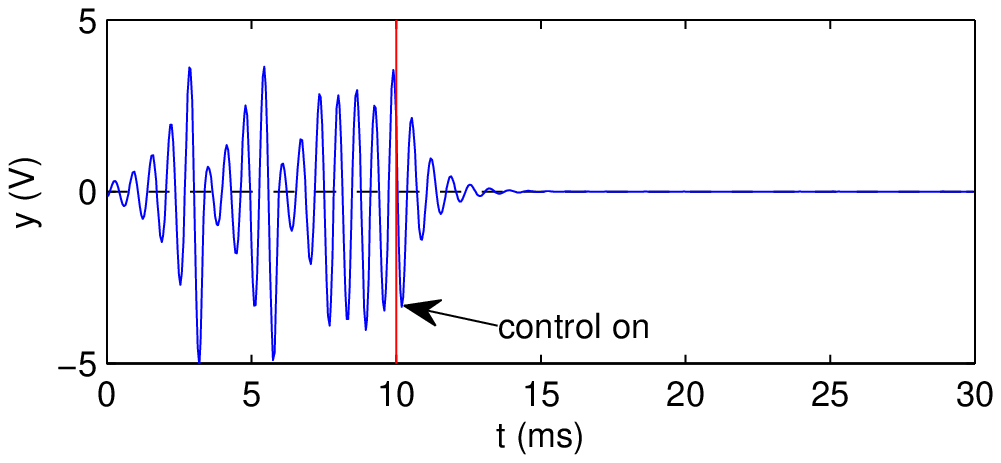}\\
b)\includegraphics[width=0.9\columnwidth]{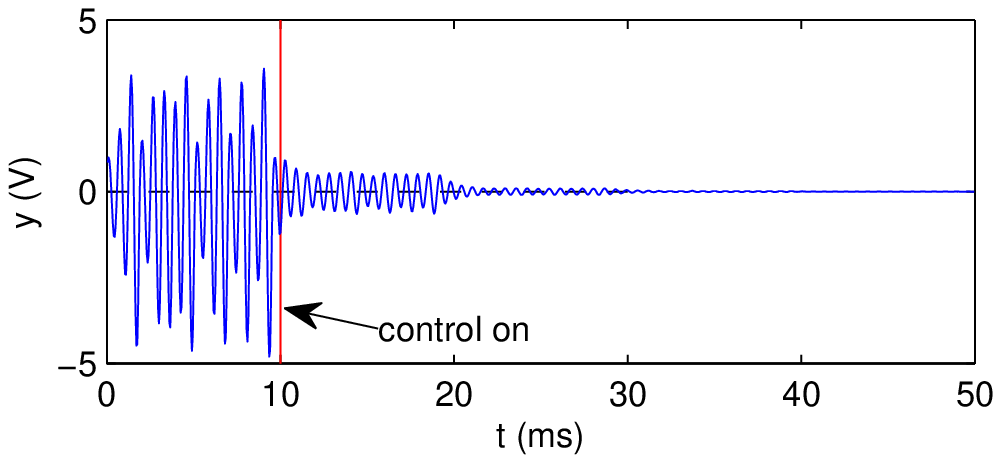}\\
c)\includegraphics[width=0.9\columnwidth]{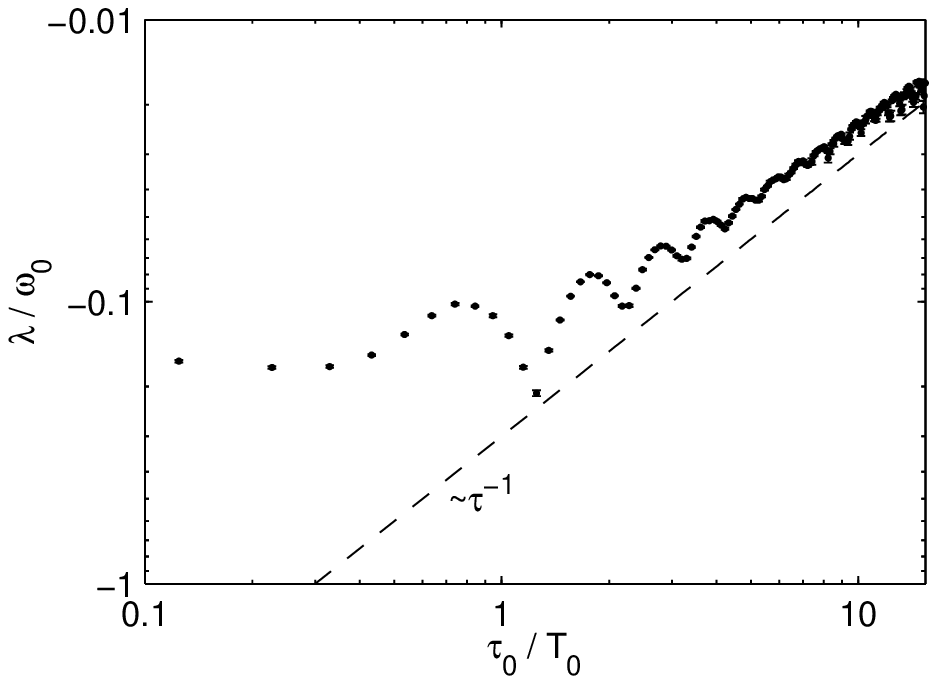}
\caption{Analysis of transients for low-pass filtered feedback control. a) Transient at $\tau_0=T_0$. b) Transient at $\tau_0=14T_0$. c) Power law dependence of the stability exponent $\lambda$ on the delay $\tau_0$.}
\label{transients}
\end{figure}

The robustness of the control methods is in principle already proven, because they withstand the intrinsic imperfections of our experiment. To further study this feature, we repeat selected measurements in the presence of additive noise. To this end the signal of a noise generator is injected into the $y$-component of the circuit. The power spectrum of the generator covers a frequency band from $f_{low}=20$Hz to $f_{high}=40$kHz (-3dB bandwidth), so it can be modelled by white noise. In the range, which is relevant for the circuit, the spectral power density is $S(f)=10\text{V}^2/$MHz with an accuracy of $\pm1$dB. The equation of motion for the $y$-component is then changed to
\begin{equation}
\dot{y}=\omega_0\left(x+(a-c)y+k(y_\tau-y)+\xi(t)\right)
\end{equation}
with
\begin{equation}
\langle\xi(t)\xi(t')\rangle=2D\delta(t-t')\;.
\end{equation}
The signal is injected via a $100\text{k}\Omega$ resistor instead of the standard $R_4=10\text{k}\Omega$, so that the noise intensity is rescaled by a factor 100. Thus we obtain $2D=100(\text{mV})^2/$Hz. We perform a parameter scan for each control method, in which the noise-induced oscillations in terms of the standard deviation $\sigma_y$ of $y(t)$ are measured. A low value corresponds to a high degree of stability and vice versa. In order to differentiate between very low values of $\sigma_y$ we evaluate the quantity
\begin{equation}
\eta=\frac{D}{2\sigma_y^2}\;.
\label{etadef}
\end{equation}
The choice of the definition for $\eta$ is motivated by the application of a fictive low-pass filter. If a simple low-pass filter - like the one we use in our previous investigation - with a decay constant $\eta$ was excited by white noise with spectral power density $2D$, it would in average deviate from its rest state by an amount $\sigma=\sqrt{D/2\eta}$, which can be calculated using standard techniques. By analogy to this analytically treatable case we evaluate the decay constant $\eta$. The results are shown in Fig.~\ref{noisyscans}.
For all control techniques there is no visible loss of robustness at large delay times. This result is even more peculiar, since from numerical simulations we have seen, that the maximum exponent of the controlled system tends towards zero for large delays, see Sec.~\ref{mechanism}. We also investigated this issue in experiment. By switching on the control force $k$ at some time $t_0$ one can observe transients and determine the relaxation rate towards the stable fixed point. This has been done for the low-pass filter control method at different delay times. Fig.~\ref{transients} shows some typical transients, which appear after the onset of control for small and large delay times $\tau_0$. The evaluation of the decay rate $\lambda(\tau_0)$ for all transients also reveals a power law, which would be typical in the presence of a pseudocontinous spectrum, see e.g.~\cite{Yanchuk:09}. In summary, these findings clearly show, that the presented control methods work well for large delay times, although the leading exponent is of order $\mathcal{O}(1/\tau_0)$, which at first sight is counterintuitive. But the reaction to noise takes into account more of the stable modes, which then leads to the $\tau_0$-independent behavior.

\section{Stabilization of periodic orbits}
\label{orbits}

Variable or distributed delay feedback is also suitable for the control of unstable periodic orbits. In contrast to the control of fixed points, here one cannot apply an arbitrary modulation, because delays which differ from multiples of the orbit period will lead to an invasive feedback force. Any delay modulation is allowed, which in the limit of high modulation frequency creates a distribution
\begin{equation}
\rho(\theta)=\sum_{n=1}^\infty a_n\delta(\theta-nT_p)\;,
\end{equation}
where $T_p$ is the period of the orbit and $\sum_n a_n=1$. Note that in this general notation also the case of extended TDFC is included \cite{Socolar:94,Pyragas:95,Bleich:96}. For ETDFC the coefficients have to be chosen as $a_n=(1-R)R^{n-1}$, where $R$ is the recursion factor of the geometric sum. $R\rightarrow0$ gives $a_n=\delta_{n,1}$, which corresponds to TDFC. A value of $R$ close to $1$ or $-1$ takes account of many multiples of the delay time.

\subsection{Two delay control}

\begin{figure}
\includegraphics[width=\columnwidth]{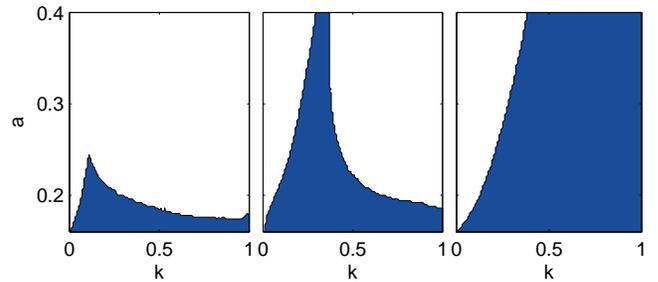}
\caption{Scan of the $a$-$k$-parameter plane for control of period-1 orbit. Control domains (standard deviation of $y(t-T_p)-y(t)$ is $<0.1$V) are marked in blue. Left: TDFC. Middle: ETDFC with $R=0.5$. Right: Two-delay distributed delay feedback.}
\label{akscans}
\end{figure}

In this paper we want to focus on those types of modulation (distribution), which create contributions to the sum in a not very large interval of delay times. The simplest case is given by the superposition of two delays similar to Eq.~(\ref{twodeleq}). To show the theoretically predicted performance improvement in experiment, we choose to control the unstable period-1 orbit of our chaotic circuit, which has a period $T_p\approx T_0=2\pi/\omega_0$. The instability of this orbit increases monotonically with the main control parameter $a$, so we study the stabilization of the orbit in the $k$-$a$ parameter plane. Analog to the automatic scan technique used for the fixed point we measure the deviation between $y(t)$ and $y(t-T_p)$ to determine whether the orbit is successfully controlled. We want to compare our two-delay control technique to ordinary TDFC in a regime where the classical method largely fails. This can be achieved by setting a large delay time, because an orbit can only be stabilized if $\tau<2/\Re\{\Lambda_0\}$, where $\Lambda_0$ is the floquet exponent of the uncontrolled orbit. The claim for noninvasiveness and the torsion of the orbit restrict the possible values of $\tau$ to odd multiples of $T_p$. So we set for TDFC a delay time $\tau=3T_p$, for which the general limit of stability is more easily exceeded than for the standard setting $\tau=T_p$. As a suitable example of distributed delay control we take a superposition of two different delayed signals with delays $\tau_1=3T_p$ and $\tau_2=4T_p$. The results are shown in Fig.~\ref{akscans} and clearly demonstrate the improvement of stabilization. Additionally, we repeat the experiment using the extended TDFC method. Here a single delay line is also set to $\tau=3T_p$, as for the TDFC measurement. But instead of receiving only the direct signal $y(t)$, the delay line receives a sum of $y(t)$ and its own output, which leads to the geometric sum of multiple delays. Direct input and recursive input are equally weighted, so we obtain a recursion factor $R=0.5$. The resulting control domain is included in Fig.~\ref{akscans} and demonstrates the improvement compared to simple TDFC. However, the control interval narrows for increasing instability, which in the examined range was not observed for the two-delay control method.

\subsection{Finite modulation frequency}

\begin{figure}
\includegraphics[width=\columnwidth]{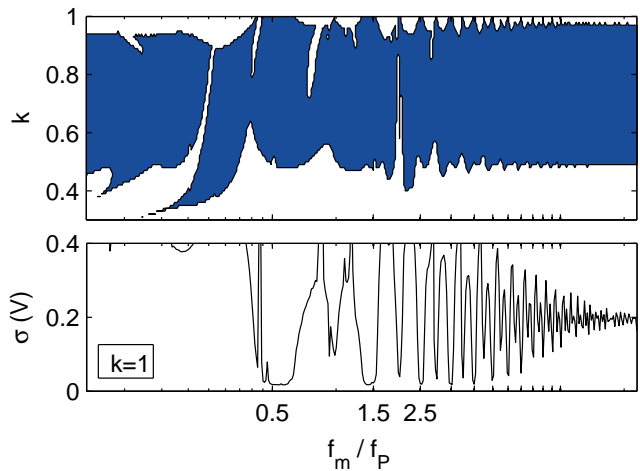}
\caption{Control of P1 orbit at $a=0.5$ with $f_p=T_p^{-1}=1.51\mathrm{kHz}$. Delay modulation between $\tau_1=T_p$ and $\tau_2=2T_p$ with frequency $f_m$. Top: Stabilization domain (blue, $\sigma<0.1$V) in dependence of feedback gain $k$ and modulation frequency. Bottom: Strength $\sigma$ of remaining control signal at $k=1$ showing optimal control at odd multiples of $f_p/2$. Displayed frequency range: 100Hz-40kHz.}
\label{fmodscan}
\end{figure}
The direct superposition of two delays as presented above represents the distributed delay limit of fast switching between the delays $\tau_1$ and $\tau_2$. Beyond this we aim to address two further questions. 1) Which modulation frequencies in variable-delay control are high enough to obtain the same results as for distributed delays? 2) It has been shown theoretically that a special choice of the modulation frequency leads to a further improvement compared to distributed delays. Is this phenomenon observable in experiment?

To clarify this, we construct a setup similar to the previous one, but with the possibility to modulate the weights of each delay line using a variable gain. The gain is implemented by multipliers from type AD633. We could also have built a setup, in which we switch between the delay lines. But it turned out, that at the moment of switching high frequency distortions are generated, which spoil the noninvasiveness of the control and thus make it hard to determine, when the orbit has been controlled successfully. In order to avoid distortions arising from very high frequency components, we modulate between the delay lines sinusoidally, so that the control force becomes
\begin{equation}
k\left(\mu(t)y_{\tau.1}+(1-\mu(t))y_{\tau.2}-y\right)\;,
\end{equation}
where $\mu(t)=\tfrac{1}{2}(1+\sin(2\pi f_m t))$ is the modulation signal. We are interested in the frequency dependence of the control boundaries $k_{\min}$ and $k_{\max}$. Without considerable distortions we can only apply a coupling gain $k\in[0,1]$. But similar to the case of two-delay control as shown in Fig.~\ref{akscans} we come across the problem that $k_{\max}>1$ for standard parameters. However, an increase of the parameter $a$ in our circuit leads to orbits with a higher degree of instability, for which the control interval should become narrow enough to be observed completely. At least for $a=0.5$ we found the control boundaries of the period-1 orbit to lie in the detectable range. For this orbit we measure the strength $\sigma(k,f_m)$ of the control signal before gain, which vanishes in case of successful stabilization. The detected control domain for modulation between $\tau_1=1T_p$ and $\tau_2=2T_p$  is shown in Fig.~\ref{fmodscan}.

In general, one can see that the dependence on a finite modulation frequency is not very strong, meaning that - as found for the fixed point - variable delay works about as well as the corresponding distributed delay. The transition to the distributed delay limit can be seen in the regime about one order of magnitude above the main oscillation frequency $f_p$. Below, there are interactions between delay modulation and frequency components of the oscillator, which in their complete appearance are nontrivial. The only phenomenon, which already has been theoretically explained, is the enhancement of stabilization at half the orbit frequency and odd multiples of it. The control domain is extended towards larger values of $k$, which confirms this prediction very well.

\section{Mechanism of stabilization}
\label{mechanism}

Stability analysis for the presented modulated (distributed) feedback technique can be regarded as mostly covered by our previous theoretical work as well as others \cite{Atay:03,Ahlborn:04,Ahlborn:06,Konishi:10,Kyrychko:11}. For completeness we give here an interpretation as well as exemplifying analytical and numerical calculations for the mechanism behind the observed phenomena. Essentially it can be discussed in terms of a normal form
\begin{equation}
\dot{z}(t)=(\alpha_0+i\omega_0)z(t)+\kappa\left(z(t-\tau(t))-z(t)\right)
\end{equation}
with $\alpha_0>0$ and $\omega_0,\kappa\in\mathbb{R}$. In the distributed-delay limit this equation turns to
\begin{equation}
\dot{z}(t)=(\alpha_0+i\omega_0)z(t)+\kappa\left(\int_0^\infty\rho(\theta)z(t-\theta)d\theta-z(t)\right)\;.
\end{equation}
With an exponential ansatz we obtain the eigenvalue equation
\begin{equation}
\lambda=\alpha_0+i\omega_0+\kappa\left(\int_0^\infty\rho(\theta)\e^{-\lambda\theta}d\theta-1\right)\;.
\end{equation}
If $\rho(\theta)$ is nonzero only between $\tau_0-\varepsilon$ and $\tau_0+\varepsilon$, then we can refine the above expression further to
\begin{equation}
\begin{split}
\lambda&=\alpha_0+i\omega_0+\kappa\left(\e^{-\lambda\tau_0}\int_{-\varepsilon}^{+\varepsilon}\rho(\tau_0+\theta)\e^{-\lambda\theta}d\theta-1\right)\\
&=\alpha_0+i\omega_0+\kappa\left(\e^{-\lambda\tau_0}\chi(\lambda,\varepsilon)-1\right)\;.
\end{split}
\label{chiexpression}
\end{equation}
Here the quantity $\chi(\lambda,\varepsilon)$ summarizes the effect of a given modulation (distribution). In the non-modulated case we have simply $\chi(\lambda,0)\equiv 1$, which then reveals the known characteristic equation for TDFC. Given Eq.~(\ref{chiexpression}) it is easy to imagine a scenario, in which improved stabilization is obvious: If one were able to find a distribution $\rho(\theta)$ such that $\chi(\lambda,\varepsilon)=0$, then Eq.~(\ref{chiexpression}) would reduce to
\begin{equation}
\lambda=\alpha_0+i\omega_0-\kappa\;.
\end{equation}
Here for $\kappa>\alpha_0$ the fixed point would be stable, regardless of how large the delay time is. In practice it is difficult to keep $\chi\equiv 0$, but the effect already appears to be strong enough if $\chi$ is only close to zero. By the following examples we show, how the presented modulation (distribution) techniques satisfy this criterion.

\subsection{Numerical examples}

\paragraph{Delay modulation}

\begin{figure}
\includegraphics[width=\columnwidth]{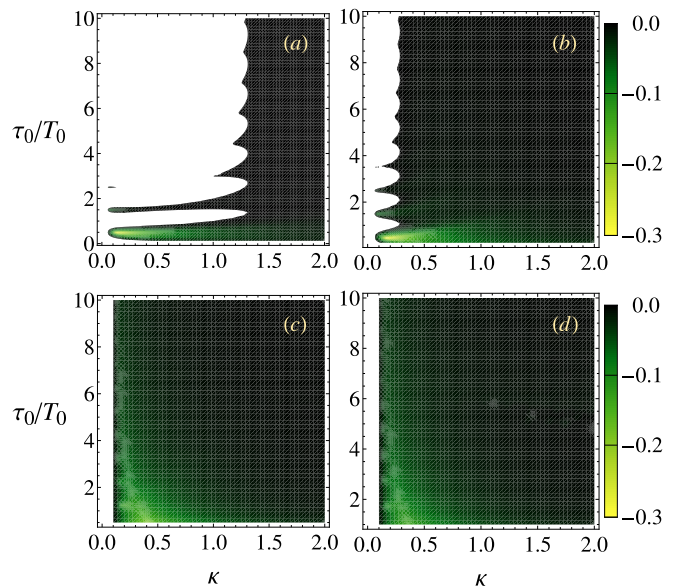}
\caption{Stability domains of the fixed point $z=0$ in the plane parametrized by the feedback gain $\kappa$ and the nominal time-delay $\tau_0/T_0$ for a triangular modulation of the delay and different modulation amplitudes $\varepsilon/T_0$: (a) $\varepsilon/T_0=0.125$, (b) $\varepsilon/T_0=0.25$, (c) $\varepsilon/T_0=0.5$, (d) $\varepsilon/T_0=1$. White: no control. Colors correspond to the real part of $\lambda$ after Eq.~(\ref{char1}).}
\label{vdfc-1}
\end{figure}

Fast triangular modulation between two delays $\tau_1=\tau_0-\varepsilon$ and $\tau_2=\tau_0+\varepsilon$ is equivalent to a delay distribution
\begin{equation}
\rho(\theta)=
\begin{cases}
1/(2\varepsilon) &,\; \tau_1\le\theta\le\tau_2,\\
0 & \text{, else}\;.
\end{cases}
\end{equation}
This leads to
\begin{equation}
\chi(\lambda,\varepsilon)=\frac{\sinh(\lambda\varepsilon)}{\lambda\varepsilon}\;.
\end{equation}
Eq.~(\ref{chiexpression}) then becomes
\begin{equation}
\lambda-\alpha_0-i\omega_0
+\kappa\left(1-\e^{-\lambda\tau_0}\sinh(\lambda\varepsilon)/(\lambda\varepsilon)\right)=0\;.
\label{char1}
\end{equation}

\begin{figure}
\includegraphics[width=\columnwidth]{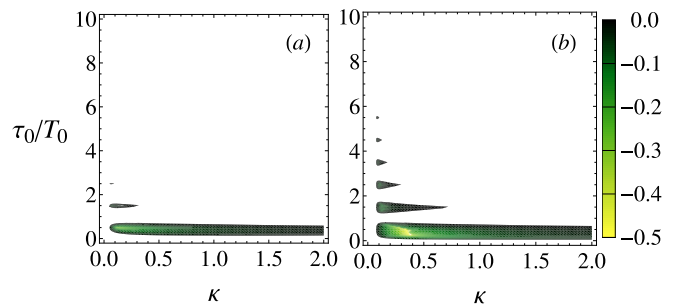}
\caption{Stability domains in the absence of delay modulation: (a) Pyragas' time-delayed feedback control (TDFC), (b) extended time-delayed feedback control (ETDFC) due to Socolar et al. \cite{Socolar:94} with $R=0.5$. White: no control. Colors correspond to the real part of $\lambda$ from Eq.~(\ref{chiexpression}).}
\label{etdfc}
\end{figure}
Numerical calculation of the stability domain from the characteristic Eq. (\ref{char1}) gives the diagram shown in Fig. \ref{vdfc-1}. Panels $(a)$--$(d)$ correspond to different values of the amplitude parameter $\varepsilon$ normalized by the intrinsic period $T_0=2\pi$ of the uncontrolled system: $(a)$ $\varepsilon/T_0=0.125$, $(b)$ $\varepsilon/T_0=0.25$, $(c)$ $\varepsilon/T_0=0.5$, $(d)$ $\varepsilon/T_0=1$. The parameters of the unstable spiral are $\alpha_0=0.1$ and $\omega_0=1$. The color-codes at the right end of the panel denote the values of the largest real part of the complex eigenvalues obtained numerically from Eq. (\ref{char1}). The shaded region in each panel denotes the negative values of the maximum  real part of the complex eigenvalues, i.e. it depicts only the control parameters for which the stabilization is successful. 
As it is seen from panels $(a)$--$(d)$, increase of the modulation amplitude $\varepsilon/T_0$ leads to a larger control domain. Further numerical analysis show that in this case the increase of the stability area is monotonic upon increasing the value of $\varepsilon/T_0$. To illustrate the superiority of the control method with respect to the standard delayed feedback control schemes,
in Fig. \ref{etdfc} we show numerically calculated control domains in the case of TDFC (panel a) and ETDFC (panel b) for the same parameters of the uncontrolled system. In the case of ETDFC, the memory parameter is chosen as $R=0.5$.

\paragraph{Two delays}

\begin{figure}
\includegraphics[width=\columnwidth]{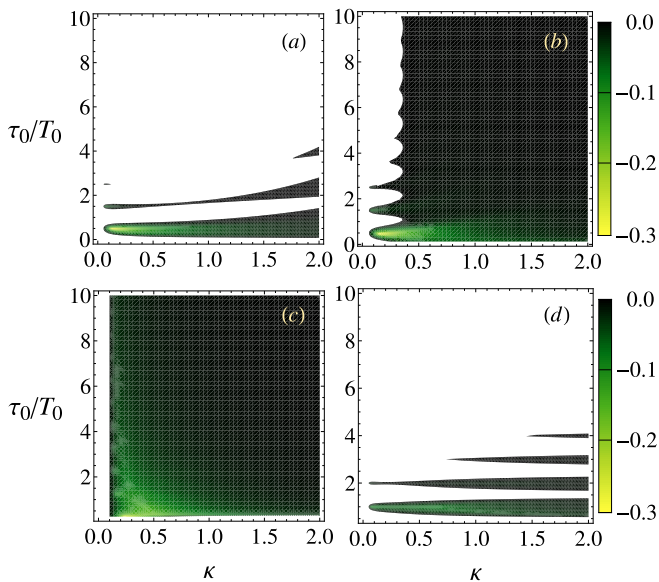}
\caption{Stability domains for the controlled origin in the parametric plane spanned by the feedback gain $\kappa$ and the nominal delay $\tau_0/T_0$ for a square-wave modulation of the delay. The modulation amplitudes are: (a) $\varepsilon/T_0=0.0625$, (b) $\varepsilon/T_0=0.125$, (c) $\varepsilon/T_0=0.25$, (d) $\varepsilon/T_0=0.5$. White: no control. Colors correspond to the real part of $\lambda$ after Eq.~(\ref{char2}).}
\label{vdfc-2}
\end{figure}

Fast and symmetric switching between two delays $\tau_1=\tau_0-\varepsilon$ and $\tau_2=\tau_0+\varepsilon$ (i.e. square-wave modulation) is equivalent to a delay distribution
\begin{equation}
\rho(\theta)=\tfrac{1}{2}\left(\delta(\theta-\tau_0+\varepsilon)+\delta(\theta-\tau_0-\varepsilon)\right)\;.
\end{equation}
According to Eq.~(\ref{chiexpression}) we obtain
\begin{equation}
\chi(\lambda,\varepsilon)=\cosh(\lambda\varepsilon)\;.
\end{equation}
In this case we arrive at the quasipolinomial equation
\begin{equation}
\lambda-\alpha_0-i\omega_0+\kappa\left(1-\e^{-\lambda\tau_0}\cosh(\lambda\varepsilon)\right)=0\;.
\label{char2}
\end{equation}
In Fig. \ref{vdfc-2} we give the numerically calculated control domains by using Eq. (\ref{char2}). Contrary to the observations for the domain enlargement in the case of a triangular-wave modulation, here the variation of the stability area is non-monotonic upon increasing the modulation amplitude. The control domain increases its area until a certain value of the modulation amplitude, after which the area is decreased, and the stability islands are rearranged. With further increase of the amplitude $\varepsilon$, the stability area is increasing again, and this alternating behavior of increasing and decreasing the area of stabilization continues. This effect could also be reproduced in experiment. In contrast, for the modulation case with increasing modulation amplitude the control performance gets worse in experiment, contrary to the theoretical expectation. This is explained by the fact, that in our modulation setup the modulation period $T_m$ is also proportional to the modulation amplitude $\varepsilon$ by construction. So the failure of control for large values of $\varepsilon$ can be related to the low modulation frequency. The numerical results obtained by the distributed-delay approximation are no longer valid in this case.

\paragraph*{Interpretation}

From the above examples we get an understanding of the basic mechanism behind variable (distributed) delay feedback control. If we have a closer look at the boundaries of control, in which $\Re\{\lambda\}=0$, we see that the coefficient $\chi$ reduces to
\begin{equation}
\chi_a=\frac{\sin(\omega\varepsilon)}{\omega\varepsilon}
\end{equation}
for the constant distribution in example (a), and
\begin{equation}
\chi_b=\cos(\omega\varepsilon)
\end{equation}
for the two-peak distribution in example (b). These are the Fourier transforms of the corresponding delay distributions, which can also be seen from the generating eq.~(\ref{chiexpression}) if $\lambda$ is set to $i\omega$. So at the threshold to stability the distribution in (a) acts like a `single slit' and the distribution in (b) acts like a `double slit', respectively. Optimal control can in these terms be found for those values of $\varepsilon$, which create complete destructive `interference', i.e, $\chi=0$. Numerical calculations for our setup show, that $\omega\approx\omega_0=2\pi/T_0$ is a valid approximation, so that we find optimal control at $\varepsilon=nT_0/2$ for example (a), and $\varepsilon=(n+1/2)T_0/2$ for example (b), which is in excellent agreement with the experimental findings.

\subsection{Alternative interpretaton for periodic orbits}

\begin{figure}
a)\includegraphics[width=0.85\columnwidth]{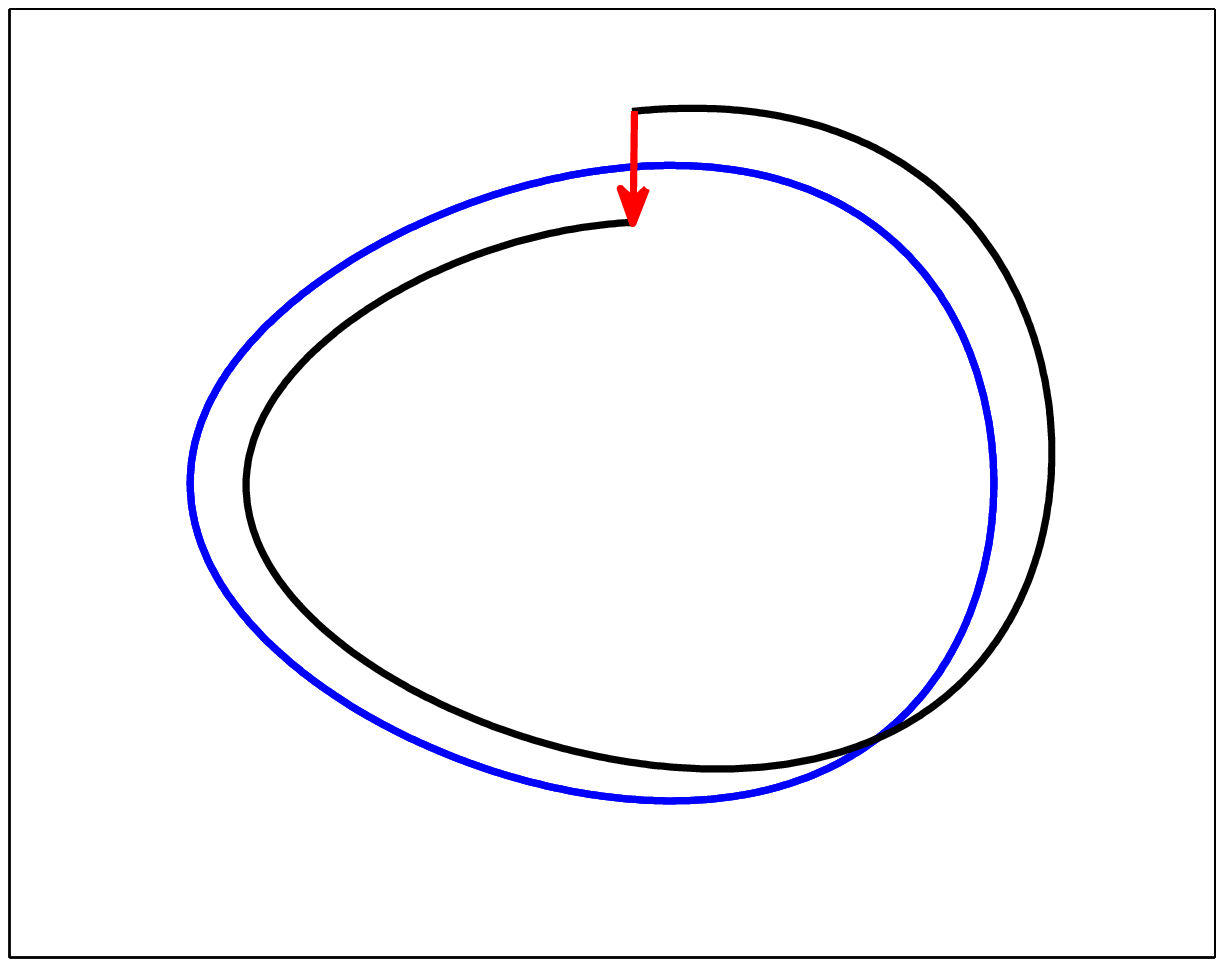}\\
b)\includegraphics[width=0.85\columnwidth]{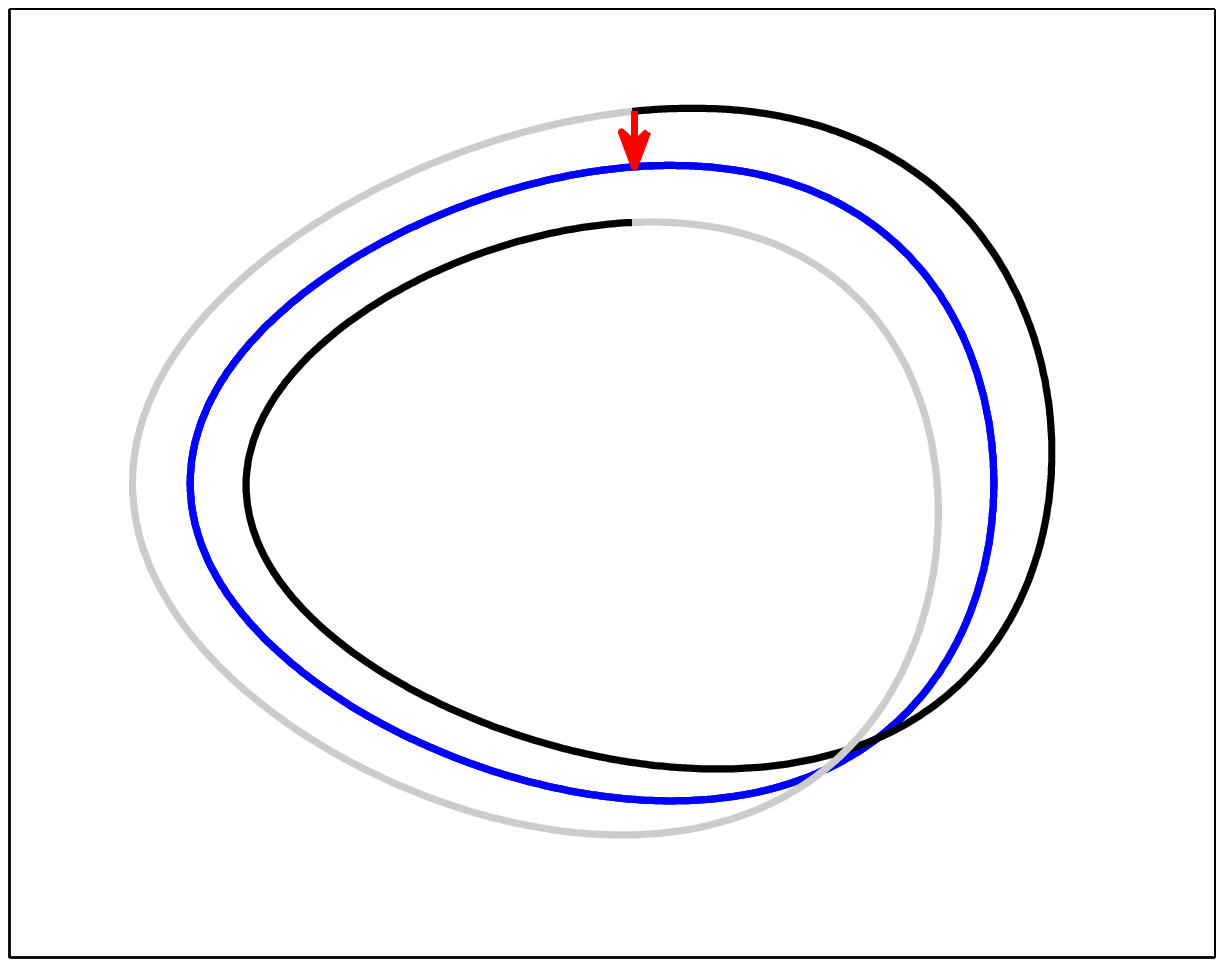}
\caption{Illustration of trajectories close to an unstable periodic orbit (blue) in the presence of a) time-delayed feedback control b) two-delay feedback control. Blue: target orbit, red: difference used for control force at $t=t_0$, black: neighborhood trajectory for $t\in[t_0-\tau,t_0]$, gray: neighborhood trajectory for $t\in[t_0-2\tau,t_0-\tau]$.}
\label{orbitfigs}
\end{figure}

Let $\boldsymbol{\xi}(t)$ be an unstable trajectory of an arbitrary dynamical system with
\begin{equation}
\dot{\boldsymbol{\xi}}(t)=\mathbf{f}(\boldsymbol{\xi}(t))\;.
\end{equation}
Regardless of what kind of solution this is - a chaotic, a periodic or a fixed point - applying the control force
\begin{equation}
\dot{\mathbf{x}}(t)=\mathbf{f}(\mathbf{x}(t))+k\left(\boldsymbol{\xi}(t)-\mathbf{x}(t)\right)
\label{optcontrol}
\end{equation}
will always result in stabilization of $\boldsymbol{\xi}(t)$ for $k>k_c$. The disadvantage of this concept is well-known. One has to know the exact solution $\boldsymbol{\xi}(t)$ and have access to all degrees of freedom in $\mathbf{x}(t)$. Most of the established control methods are based on finding a proper approximation of $\boldsymbol{\xi}(t)$ to create a difference control force close to the optimal force in Eq.~(\ref{optcontrol}). In the case of TDFC the delayed vector $\mathbf{x}(t-\tau)$ is supposed to approximate a periodic orbit with period $\tau$, but the introduced delay term gives rise to instabilities for some parameter constellations. In other words, the approximation error between $\mathbf{x}(t-\tau)$ and the target state $\boldsymbol{\xi}(t)$ may result in failure of the method. For TDFC it is well-known, that stabilization is not successful, if either $k$ or $\tau$ is too large, depending on the stability properties of the uncontrolled orbit. In control theory such an issue would be explained in terms of \textit{overcompensation}, which means that the control force first drives the trajectory towards the desired orbit, but then shoots over the target resulting in an increasing oscillation. Fig.~\ref{orbitfigs}a shows how in TDFC the control force of a flip-orbit is created at the threshold to stabilization, i.e. for $\Re\{\Lambda\}=0$ and $\Im\{\Lambda\}=\pi/T_p$, indicating the possibility of an overshoot. For our control method using a superposition of two delays at $\tau_1=T_p$ and $\tau_2=2T_p$ the situation is different. Constructing again the situation at the threshold to stabilization we show in Fig.~\ref{orbitfigs}b that the reference signal approximates the target orbit much better than for simple TDFC. This should prevent overshoots and result in stabilization at a larger set of coupling parameters. Of course this example is just an intuitive explanation, which holds best if the neighborhood of the orbit undergoes a torsion of an odd multiple of $\pi$ within one period, even under the presence of the control force, which is in general not the case. The variable (distributed) delay feedback is still not able to control \textit{any} orbit, but its improved performance compared to simple TDFC may be helpful in many control applications.

\section{Conclusions}
\label{conclusions}

We have shown that the use of variable (distributed) delays in a time-delayed feedback control scheme can significantly improve the control performance for unstable fixed points as well as for unstable periodic orbits. We have proven the robustness and reliability of our control method by an experiment with a chaotic electronic oscillator and digital delay lines. The autonomous diode oscillator has an unstable fixed point, which is susceptible for control by the Pyragas method and its modifications, as well as a set of periodic orbits, which are amenable to control by the same methods.\\

The fixed point allows for any delay modulation (distribution), from which we realized three different types: 1) modulation of the delay time via modulation of the clock frequency driving the delay lines, 2) an exponential delay distribution using one delay line and a low-pass filter, and 3) a superposition of two equally weighted delays using two parallel delay lines. The effect of all variations is comparable: New domains of control are generated at large values of delay time and coupling strength. For an optimal value of the modulation amplitude (distribution width) we obtain one connected domain of control, which extends to arbitrarily large delay times. We were able to show that for large mean delay, control works in a robust way in spite of the decreasing magnitude of the leading stability exponent.\\

Periodic orbits allow for such a delay modulation (distribution), in which only multiples of the orbit period are included. Here we were able to show, that switching between two delays with a difference of a single period of the orbit leads to similar extensions of the control domain as observed for the fixed point. Further we were able to stabilize orbits, which cannot be stabilized by TDFC. As for the fixed point, a finite modulation frequency does not disturb the mechanism. In contrast, if the modulation frequency matches odd multiples of half the orbit frequency, we found the stabilization to be optimal.\\

The mechanism of our control method can be explained in terms of destructive interference. The intrinsic torsion of an orbit or fixed point, which is necessary for TDFC to work in principle, is used by the delay modulation (distribution) in a way, that different phases of the torsion cancel each other out. This leads to minimization of the effect of the delay term, which in most cases is an undesirable source of instability, especially at large delay times.

\bibliographystyle{apsrev}
\bibliography{References}

\begin{thebibliography}{18}
\expandafter\ifx\csname natexlab\endcsname\relax\def\natexlab#1{#1}\fi
\expandafter\ifx\csname bibnamefont\endcsname\relax
  \def\bibnamefont#1{#1}\fi
\expandafter\ifx\csname bibfnamefont\endcsname\relax
  \def\bibfnamefont#1{#1}\fi
\expandafter\ifx\csname citenamefont\endcsname\relax
  \def\citenamefont#1{#1}\fi
\expandafter\ifx\csname url\endcsname\relax
  \def\url#1{\texttt{#1}}\fi
\expandafter\ifx\csname urlprefix\endcsname\relax\def\urlprefix{URL }\fi
\providecommand{\bibinfo}[2]{#2}
\providecommand{\eprint}[2][]{\url{#2}}

\bibitem[{\citenamefont{Pyragas}(1992)}]{Pyragas:92}
\bibinfo{author}{\bibfnamefont{K.}~\bibnamefont{Pyragas}},
  \bibinfo{journal}{Physics Letters A} \textbf{\bibinfo{volume}{170}},
  \bibinfo{pages}{421} (\bibinfo{year}{1992}).

\bibitem[{\citenamefont{Schuster and Sch\"{o}ll}(2007)}]{Schuster:07}
\bibinfo{editor}{\bibfnamefont{H.~G.} \bibnamefont{Schuster}} \bibnamefont{and}
  \bibinfo{editor}{\bibfnamefont{E.}~\bibnamefont{Sch\"{o}ll}}, eds.,
  \emph{\bibinfo{title}{Handbook of Chaos Control}} (\bibinfo{publisher}{Wiley
  VCH}, \bibinfo{year}{2007}), \bibinfo{edition}{2nd} ed.

\bibitem[{\citenamefont{Socolar et~al.}(1994)\citenamefont{Socolar, Sukow, and
  Gauthier}}]{Socolar:94}
\bibinfo{author}{\bibfnamefont{J.~E.~S.} \bibnamefont{Socolar}},
  \bibinfo{author}{\bibfnamefont{D.~W.} \bibnamefont{Sukow}}, \bibnamefont{and}
  \bibinfo{author}{\bibfnamefont{D.~J.} \bibnamefont{Gauthier}},
  \bibinfo{journal}{Phys. Rev. E} \textbf{\bibinfo{volume}{50}},
  \bibinfo{pages}{3245} (\bibinfo{year}{1994}).

\bibitem[{\citenamefont{Socolar and Gauthier}(1998)}]{Socolar:98}
\bibinfo{author}{\bibfnamefont{J.~E.~S.} \bibnamefont{Socolar}}
  \bibnamefont{and} \bibinfo{author}{\bibfnamefont{D.~J.}
  \bibnamefont{Gauthier}}, \bibinfo{journal}{Phys. Rev. E}
  \textbf{\bibinfo{volume}{57}}, \bibinfo{pages}{6589} (\bibinfo{year}{1998}).

\bibitem[{\citenamefont{Ahlborn and Parlitz}(2004)}]{Ahlborn:04}
\bibinfo{author}{\bibfnamefont{A.}~\bibnamefont{Ahlborn}} \bibnamefont{and}
  \bibinfo{author}{\bibfnamefont{U.}~\bibnamefont{Parlitz}},
  \bibinfo{journal}{Phys. Rev. Lett.} \textbf{\bibinfo{volume}{93}},
  \bibinfo{pages}{264101} (\bibinfo{year}{2004}).

\bibitem[{\citenamefont{Gjurchinovski and Urumov}(2008)}]{Gjurchinovski:08}
\bibinfo{author}{\bibfnamefont{A.}~\bibnamefont{Gjurchinovski}}
  \bibnamefont{and} \bibinfo{author}{\bibfnamefont{V.}~\bibnamefont{Urumov}},
  \bibinfo{journal}{EPL (Europhysics Letters)} \textbf{\bibinfo{volume}{84}},
  \bibinfo{pages}{40013} (\bibinfo{year}{2008}).

\bibitem[{\citenamefont{Atay}(2003)}]{Atay:03}
\bibinfo{author}{\bibfnamefont{F.~M.} \bibnamefont{Atay}},
  \bibinfo{journal}{Phys. Rev. Lett.} \textbf{\bibinfo{volume}{91}},
  \bibinfo{pages}{094101} (\bibinfo{year}{2003}).

\bibitem[{\citenamefont{Konishi et~al.}(2010)\citenamefont{Konishi, Kokame, and
  Hara}}]{Konishi:10}
\bibinfo{author}{\bibfnamefont{K.}~\bibnamefont{Konishi}},
  \bibinfo{author}{\bibfnamefont{H.}~\bibnamefont{Kokame}}, \bibnamefont{and}
  \bibinfo{author}{\bibfnamefont{N.}~\bibnamefont{Hara}},
  \bibinfo{journal}{Physics Letters A} \textbf{\bibinfo{volume}{374}},
  \bibinfo{pages}{733} (\bibinfo{year}{2010}).

\bibitem[{\citenamefont{Kyrychko et~al.}(2011)\citenamefont{Kyrychko, Blyuss,
  and Sch\"oll}}]{Kyrychko:11}
\bibinfo{author}{\bibfnamefont{Y.}~\bibnamefont{Kyrychko}},
  \bibinfo{author}{\bibfnamefont{K.}~\bibnamefont{Blyuss}}, \bibnamefont{and}
  \bibinfo{author}{\bibfnamefont{E.}~\bibnamefont{Sch\"oll}},
  \bibinfo{journal}{Eur. Phys. J. B}  (\bibinfo{year}{2011}).

\bibitem[{\citenamefont{J\"{u}ngling}(2010)}]{Juengling:10}
\bibinfo{author}{\bibfnamefont{T.}~\bibnamefont{J\"{u}ngling}}, Ph.D. thesis,
  \bibinfo{school}{TU Darmstadt} (\bibinfo{year}{2010}).

\bibitem[{\citenamefont{J\"ungling et~al.}(2011)\citenamefont{J\"ungling,
  Benner, Shirahama, and Fukushima}}]{Juengling:11}
\bibinfo{author}{\bibfnamefont{T.}~\bibnamefont{J\"ungling}},
  \bibinfo{author}{\bibfnamefont{H.}~\bibnamefont{Benner}},
  \bibinfo{author}{\bibfnamefont{H.}~\bibnamefont{Shirahama}},
  \bibnamefont{and}
  \bibinfo{author}{\bibfnamefont{K.}~\bibnamefont{Fukushima}},
  \bibinfo{journal}{Phys. Rev. E} \textbf{\bibinfo{volume}{84}},
  \bibinfo{pages}{056208} (\bibinfo{year}{2011}).

\bibitem[{\citenamefont{Heiligenthal et~al.}(2011)\citenamefont{Heiligenthal,
  Dahms, Yanchuk, J\"ungling, Flunkert, Kanter, Sch\"oll, and
  Kinzel}}]{Heiligenthal:11}
\bibinfo{author}{\bibfnamefont{S.}~\bibnamefont{Heiligenthal}},
  \bibinfo{author}{\bibfnamefont{T.}~\bibnamefont{Dahms}},
  \bibinfo{author}{\bibfnamefont{S.}~\bibnamefont{Yanchuk}},
  \bibinfo{author}{\bibfnamefont{T.}~\bibnamefont{J\"ungling}},
  \bibinfo{author}{\bibfnamefont{V.}~\bibnamefont{Flunkert}},
  \bibinfo{author}{\bibfnamefont{I.}~\bibnamefont{Kanter}},
  \bibinfo{author}{\bibfnamefont{E.}~\bibnamefont{Sch\"oll}}, \bibnamefont{and}
  \bibinfo{author}{\bibfnamefont{W.}~\bibnamefont{Kinzel}},
  \bibinfo{journal}{Phys. Rev. Lett.} \textbf{\bibinfo{volume}{107}},
  \bibinfo{pages}{234102} (\bibinfo{year}{2011}).

\bibitem[{\citenamefont{Michiels et~al.}(2005)\citenamefont{Michiels,
  Van~Assche, and Niculescu}}]{Michiels:05}
\bibinfo{author}{\bibfnamefont{W.}~\bibnamefont{Michiels}},
  \bibinfo{author}{\bibfnamefont{V.}~\bibnamefont{Van~Assche}},
  \bibnamefont{and} \bibinfo{author}{\bibfnamefont{S.-I.}
  \bibnamefont{Niculescu}}, \bibinfo{journal}{IEEE Trans. Autom. Control}
  \textbf{\bibinfo{volume}{50}}, \bibinfo{pages}{493 } (\bibinfo{year}{2005}).

\bibitem[{\citenamefont{Just et~al.}(1999)\citenamefont{Just, Reibold, Benner,
  Kacperski, Fronczak, and Ho\l{}yst}}]{Just:99}
\bibinfo{author}{\bibfnamefont{W.}~\bibnamefont{Just}},
  \bibinfo{author}{\bibfnamefont{E.}~\bibnamefont{Reibold}},
  \bibinfo{author}{\bibfnamefont{H.}~\bibnamefont{Benner}},
  \bibinfo{author}{\bibfnamefont{K.}~\bibnamefont{Kacperski}},
  \bibinfo{author}{\bibfnamefont{P.}~\bibnamefont{Fronczak}}, \bibnamefont{and}
  \bibinfo{author}{\bibfnamefont{J.}~\bibnamefont{Ho\l{}yst}},
  \bibinfo{journal}{Physics Letters A} \textbf{\bibinfo{volume}{254}},
  \bibinfo{pages}{158 } (\bibinfo{year}{1999}).

\bibitem[{\citenamefont{Yanchuk and Perlikowski}(2009)}]{Yanchuk:09}
\bibinfo{author}{\bibfnamefont{S.}~\bibnamefont{Yanchuk}} \bibnamefont{and}
  \bibinfo{author}{\bibfnamefont{P.}~\bibnamefont{Perlikowski}},
  \bibinfo{journal}{Phys. Rev. E} \textbf{\bibinfo{volume}{79}},
  \bibinfo{pages}{046221} (\bibinfo{year}{2009}).

\bibitem[{\citenamefont{Pyragas}(1995)}]{Pyragas:95}
\bibinfo{author}{\bibfnamefont{K.}~\bibnamefont{Pyragas}},
  \bibinfo{journal}{Physics Letters A} \textbf{\bibinfo{volume}{206}},
  \bibinfo{pages}{323 } (\bibinfo{year}{1995}).

\bibitem[{\citenamefont{Bleich and Socolar}(1996)}]{Bleich:96}
\bibinfo{author}{\bibfnamefont{M.}~\bibnamefont{Bleich}} \bibnamefont{and}
  \bibinfo{author}{\bibfnamefont{J.}~\bibnamefont{Socolar}},
  \bibinfo{journal}{Physics Letters A} \textbf{\bibinfo{volume}{210}},
  \bibinfo{pages}{87 } (\bibinfo{year}{1996}).

\bibitem[{\citenamefont{Ahlborn and Parlitz}(2006)}]{Ahlborn:06}
\bibinfo{author}{\bibfnamefont{A.}~\bibnamefont{Ahlborn}} \bibnamefont{and}
  \bibinfo{author}{\bibfnamefont{U.}~\bibnamefont{Parlitz}},
  \bibinfo{journal}{Phys. Rev. Lett.} \textbf{\bibinfo{volume}{96}},
  \bibinfo{pages}{034102} (\bibinfo{year}{2006}).

\end{thebibliography}

\end{document}